\documentstyle[twocolumn,seceq,epsf]{jpsj}

\newcommand{\boldr}{\mbox{\boldmath $r$}}
\newcommand{\boldm}{\mbox{\boldmath $m$}}
\newcommand{\boldsig}{\mbox{\boldmath $\sigma$}}
\newcommand{\mathcomma}{\, \, \mbox{}_{,}}
\newcommand{\mathperiod}{\, \, \mbox{}_{.}}

\newcommand{\numast}{${}^{(*)}$}
\newcommand{\numdag}{${}^{(\dag)}$}
\title{Design Equation: A Novel Approach to Heteropolymer Design }
\author{ Yukito {\sc Iba}${}^1$\footnote{e-mail: iba@ism.ac.jp}, 
\ Kei {\sc Tokita} ${}^{2,3}$ and \ Macoto  
{\sc Kikuchi}${}^3$\footnote{e-mail: kikuchi@godzilla.phys.sci.osaka-u.ac.jp} }
\inst{ ${}^1$The Institute of Statistical Mathematics, \\
4-6-7 Minami-Azabu Minato-ku , Tokyo, 106-8569, Japan \\
${}^2$Department of Chemistry and Chemical Biology, \\
Harvard University, 12 Oxford Street, Cambridge MA 02138, USA \\
${}^3$Department of Physics, Osaka University, Toyonaka, 560-0043, Japan \\}

\recdate{\today}

\abst{
A novel approach to heteropolymer design is
proposed.  It is based on the criterion
by Kurosky and Deutsch, with which
the probability of a target conformation in a conformation
space is maximized at low but finite temperature.
The key feature of the proposed approach is the use
of soft spins (fuzzy monomers) that leads to a design equation, which
is an analog of the Boltzmann machine learning equation in the 
design problem. We implement an algorithm based on the design equation
for the generalized HP model 
on the $3\times3\times3$ cubic lattice and check its performance.
}

\kword{ design, heteropolymer, protein, optimization, learning,
MTP criterion, Boltzmann machine, design equation, HP model }

\begin{document}

\sloppy
\maketitle

\section{Introduction}
Recently, computer design of microscopic objects 
draws much attention of theoretical physicists.  
Designing a heteropolymer that fold  
into a given shape \cite{SG93,S95,KD95,DK96,Morrissey,SVMB96,GAS,DelKoe,Kono}
is the most challenging one because it is a key technology for protein
engineering.  This problem is also called ``inverse folding'' of
heteropolymers.  While the equilibrium conformation of a polymer with a  
given sequence of monomers
is asked in a {\it folding } problem, a sequence that
folds into a given conformation of a polymer is requested
in the corresponding {\it inverse folding} problem.

The first step in the inverse folding problem is to
formulate it as an optimization problem.
In the pioneering work \cite{SG93} of Shakhnovich and Gutin,
a sequence $\boldsig$ of monomers (or ``amino acids'') that minimize
the energy $E(\widetilde{\boldr}| \boldsig)$ of a target conformation
$\widetilde{\boldr}$ is chosen under the condition of  a fixed 
monomer composition.
This criterion  is based on the hypothesis that the energies of
misfolded conformations depend only on the composition
of the polymer.
Algorithms with this criterion have been shown to work in
many practical problems with computational efficiency.
An optimal solution, however, is not always ensured. \cite{DK96}.

Kurosky and Deutsch \cite{KD95} \cite{DK96} proposed
a different criterion. In their framework
\cite{foot1}, the equilibrium probability
\begin{equation}
\label{Gibbs}
 P_\beta(\widetilde{\boldr}| \boldsig ) =
 \frac{\exp( - \beta E(\widetilde{\boldr} | \boldsig ))}{Z_\beta}
 \mathcomma
\end{equation}
\begin{equation}
 Z_\beta= \sum_{ \boldr } \exp(-\beta E(\boldr | \boldsig ))
\end{equation}
that a polymer is found in a target conformation $\widetilde{\boldr}$
is maximized at a sufficiently low temperature $1/\beta$.
Here $\sum_{ \boldr }$ means the summation over all
possible conformations.
Hereafter, we denote this criterion as ``MTP criterion''
(Maximum Target Probability criterion).
The MTP criterion automatically excludes solutions with degenerate
ground states when there exists other solution with an unique ground state.
In addition, if we keep the temperature $1/\beta$ low but
a finite value, it selects a solution with a larger energy gap
above the ground state.  
An advantage of MTP criterion is that it is useful to 
systematically understand various algorithms as methods
for approximate maximization of  MTP criterion, for example, 
the original algorithm by Deutsch and Kurosky \cite{DK96}, the algorithm
by Morrissey and Shakhnovich \cite{Morrissey} , and the algorithms 
by Seno {\it et al.} \cite{SVMB96}. 

The purpose of this paper is to present a new
approach  to the inverse folding  problem based on the MTP criterion.
At the present status,  we are mostly concentrated on the design of
the sequences that has a given target conformation as the unique ground state.
For this purpose,  the proposed approach 
has a considerable advantage over other approaches,  e.g.,
the one in ref.~\citen{SVMB96}, which will 
be described in the section \ref{secRW} and \ref{secS}.
Although we are interested in the low temperature limit \mbox{$\beta
\rightarrow \infty$},
the use of MTP criterion at low but {\it finite} temperature is essential
in the derivation of the proposed algorithm. 

A goal of  inverse folding of hetropolymers is
to control the dynamical property of the designed sequence
as well as thermodynamic one. That is,  we want to design a sequence that 
fold fast into the desired conformation.  Around this problem, 
there are a number of issues on the relation between 
the thermodynamic property and dynamical property of 
heteropolymers below the folding temperature:
While some authors \cite{SSK,GAS} 
argue that the foldability of a heteropolymer mostly depends 
on the energy gap between the ground state and 
lower excited states, the other authors \cite{KT96} claim the relevance
of some different properties.
In our context, it will be interesting to test the dynamical properties of the sequence
designed by the proposed algorithm.  We will, however, leave it
for a future problem and restrict ourselves to an innovation of  the computational
technique in this paper. We also hope that  the development of an efficient
algorithm will contribute investigations into the above-mentioned problem.

\section{The Design Equation Approach}

Let us discuss the proposed approach -- the design equation approach.
Our starting point is the analogy to the learning in artificial neural networks.
In fact, the maximization of the probability eq.~(\ref{Gibbs}) is analogous to
Boltzmann machine learning \cite{HS86,SKH86}
in neural network theories
(and also maximum likelihood estimation in statistics
\cite{foot4}),
where the parameters that maximize the probability of
a given data are selected as an optimal solution.
Although this analogy to Boltzmann 
machine learning was already mentioned in refs.~\citen{KD95} and \citen{DK96},
it has not been fully explored in the literature.
Specifically, no direct analog of
{\it Boltzmann machine learning equation}
has been discussed in connection with the inverse folding problem.
Here we further pursue the analogy between the MTP criterion
and Boltzmann machine learning and give a new approach based on a
{\it design equation}, an analog of 
Boltzmann machine learning equation.  
For this purpose, we introduce ``soft spins'' 
or ``fuzzy monomers'' as a tool of computation.

We will discuss our approach in terms of 
the generalized HP model
\cite{D85,LD89,SG90,CD91} of protein.
This model consists of a self-avoiding polymer chain on a lattice
with two types of monomers indexed by
$1$ and $-1$. As a model of protein, the 
indices $\pm 1$ indicate hydrophobic (H) and polar (P) residues, respectively. 
The interaction energy $U$ between a pair of
monomers (the contact energy) is defined as $U(1,1)=\epsilon_1$,
$U(1,-1)=U(-1,1)=\epsilon_2$, and $U(-1,-1)=\epsilon_3$.
It acts only between monomers
on nearest neighbor sites but not consecutive along the chain.
The energy $E(\boldr|\boldsig)$ of a conformation $\boldr= \{r_i\}$
($r_i$ represents the vector of coordinates of
the $i$th monomer) of a polymer with a 
sequence $\boldsig = \{\sigma_i\}$ is written as
\begin{equation}
\label{energy1}
\label{energy}
E(\boldr | \boldsig )=  \frac{1}{2} \sum_{ij}
U(\sigma_i,\sigma_j) \, \delta(|r_i-r_j|-1) \, \eta_{ij}
\end{equation}
where $\sigma_i \in \{-1,1\}$ represents the type of $i$th  
monomer. The factor $\eta_{ij} \in \{0,1\}$ takes the value 0 if and only if
monomers $i$ and $j$ are consecutive along the chain.
For later convenience, we note that the interaction energy
$U(\sigma_i,\sigma_j)$ of the model is written as
\begin{equation}
\label{U1}
U(\sigma_i,\sigma_j)=\epsilon_a \sigma_i \sigma_j +
\frac{\epsilon_b}{2} (\sigma_i+\sigma_j)+\epsilon_c
\end{equation}
\begin{equation}
\epsilon_a=\frac{\epsilon_1-2\epsilon_2+\epsilon_3}{4}, \,
\epsilon_b=\frac{\epsilon_1-\epsilon_3}{2}, \,
\epsilon_c=\frac{\epsilon_1+2\epsilon_2+\epsilon_3}{4} \mathperiod
\end{equation}

At this point, we
introduce ``soft spins'' $\boldm=\{m_i\}$ each of that takes a
continuous value $-1 \le m_i \le 1$ and substitute
the original binary variables $\boldsig = \{\sigma_i\}$.  
Non-integer values of the
variable $m_i$ (``fuzzy monomer'') have no physical meaning in the
generalized HP model. They are, however, convenient tools for the
computation, as they are
in the Hopfield-Tank method \cite{HT85} for
combinatorial optimization problems.  
We take the following form eq.~(\ref{softE}) of the energy function,
which is a straightforward extension of eq.~(\ref{energy}) to the soft spin  
model.
\begin{equation}
\label{softE}
E^{*}(\boldr | \boldm )= \frac{1}{2} \sum_{ij}
U^{*}(m_i,m_j) \, \delta(|r_i-r_j|-1) \, \eta_{ij}
\end{equation}
The only constraint imposed on the
modified energy function $E^{*}(\boldr | \boldm)$ with continuous
variables $\boldm=\{m_i\}$ is that it coincides with the original
energy function eq.~(\ref{energy}) when $|m_i|=1$ for all $i$.
In this paper, we consider two possible choices for $U^*$,
\begin{equation}
\label{Uone}
U^1(m_i,m_j)=\epsilon_a m_i m_j +
\frac{\epsilon_b}{2} (m_i+m_j)+\epsilon_c \mathcomma
\end{equation}
and
\begin{equation}
\label{Utwo}
U^2(m_i,m_j)=\epsilon_a m_i m_j +
\frac{\epsilon_b}{2} (m_i|m_i|+m_j|m_j|)+\epsilon_c \mathperiod
\end{equation}
The former expression eq.~(\ref{Uone}) is an obvious extension
of the original energy function eq.~(\ref{U1}) and it is easy to see that
the latter expression eq.~(\ref{Utwo}) is also coincides with
the original energy function when $|m_i|=|m_j|=1$.

If we substitute these energy functions in 
eq.~(\ref{Gibbs}) of the MTP criterion, it gives

\begin{equation}
\label{mGibbs}
 P_\beta (\widetilde{\boldr} | \boldm)=
 \frac{\exp(-\beta E^{*}(\widetilde{\boldr} |  \boldm))}{Z_\beta}
\mathcomma
\end{equation}
\begin{equation}
Z_\beta =\sum_{ \boldr }
\exp(-\beta E^{*}(\boldr | \boldm))
\mathperiod
\end{equation}
The variables $\{m_i\}$ that maximize the expression
eq.~(\ref{mGibbs}) take, in general, non-integer values
and not necessary satisfy the relations $-1 \le m_i \le 1$.
Then we introduce a penalty term
\begin{equation}
V_{p}(\boldm) = \frac{\lambda}{4} \sum_i ( m_i^2-1 )^2
\end{equation}
to force them towards $1$ or $-1$.  The value of the parameter
$\lambda$ controls the strength of the penalty.
Using this penalty term, we arrive at the cost function
\begin{equation}
\label{cost}
V_{cost}(\boldm)= - \log P_\beta (\widetilde{\boldr}|\boldm) + V_{p}(\boldm)
\end{equation}
to be minimized in the present approach.

The use of the soft spins and the penalty term
are key features of our approach.
By virtue of them, we can differentiate the cost function
(\ref{cost}) with
$m_i$ and write down a set of equations,
\begin{equation}
\label{leq}
\tau \frac{dm_i}{dt} = - \frac{\partial V_{cost}}{\partial m_i}=
f_i(\beta,\boldm) - \lambda m_i(m_i^2-1) \mathcomma
\end{equation}
\begin{eqnarray}
\label{learn}
f_i(\beta,\boldm)
&=&  \beta
\sum_j \frac {\partial U^*(m_i,m_j)} {\partial m_i}\nonumber\\
 \times \hspace{5mm} && \hspace*{-8mm} \{ \delta(|\widetilde{r}_i-\widetilde{r}_j|-1)-
\langle \delta(|r_i-r_j|-1) \rangle_\beta \}
\end{eqnarray}
that minimize the cost function eq.~(\ref{cost}) with a gradient decent
method. Here the variable $t$ is a fictitious time and the constant
$\tau$ controls the time scale.
The average $\langle \delta(|r_i-r_j|-1) \rangle_\beta$
indicates the canonical average of $\delta(|r_i-r_j|-1)$ at the inverse
temperature $\beta$, i.e.,
\begin{equation}
\langle \delta(|r_i-r_j|-1) \rangle_\beta
= \sum_{ \boldr } \delta(|r_i-r_j|-1)
P_\beta (\boldr | \boldm)  \mathperiod
\end{equation}
We omit the factors $\eta_{ij}$ in the expression of $f_i$ in
eq.~(\ref{learn}) because their effects cancel between the first and
second terms in the brace \{ \} in eq.~(\ref{learn}).  

In this paper, we refer to the 
set of equations, eqs.~(\ref{leq}) and (\ref{learn}),
as the {\it design equation} for this problem.
The design equation
is nothing but an analog in the inverse
folding problem of the Boltzmann machine learning equation.  When the
value of the control parameter $\lambda$ is gradually increased to
$+\infty$ as the fictitious time $t \rightarrow \infty$, the value of
each soft spin $m_i$ converges to $\pm 1$, which defines a sequence
with proper meaning in the original problem. It is easy to see that
a sequence whose unique ground state coincides with the desired  
conformation satisfies
the equation ${\partial V_{cost}}/{\partial m_i}=0$.  
Thus, for sufficiently low temperature $1/\beta$,
the output of the present procedure is a candidate for the solution
of the original problem with the MTP criterion.

Details of our  implementation of the design equation 
are shown in the followings:
\begin{enumerate}
\item Initialization. \\
 Set $k:=1$, $m_i:=m_i^0$, $\lambda:=\lambda_0$.
\item Calculation of the canonical averages.\\
Calculate $\langle \delta(|r_i-r_j|-1) \rangle_\beta$
by the exact enumeration or by a Monte Carlo procedure.
The former is possible only for short chains.
\item An iteration of the discretized design equation. \\
{\samepage
\begin{eqnarray}
& &f_i := \beta
\sum_j \frac {\partial U^*(m_i,m_j)}{\partial m_i}\nonumber\\
 & &\qquad\times  \{ \delta(|\widetilde{r}_i-\widetilde{r}_j|-1)-
\langle \delta(|r_i-r_j|-1) \rangle_\beta \}\nonumber\\
& &m_i := m_i + \Delta
\left \{ f_i - \lambda m_i(m_i^2-1)
\right \} \mathperiod\nonumber
\end{eqnarray}
}
\item Clipping.\\
If $m_i >1$, set $m_i:=1$. \\
If $m_i< -1$, set $m_i:=-1$.
\item Increment the variables and  check the convergence. \\
Set $k:= k+1$ and $\lambda:= \lambda + a$.
If a prescribed stopping criterion is satisfied or the counter $k$
exceeds a prescribed maximum of the iteration,
end up the calculation. Else, return to step 2.
\end{enumerate}
The constant $a$ controls the increase of the strength
$\lambda$ of the penalty and the constant $\Delta$ controls
the size of steps in the gradient decent.
The tunable control parameters are $\lambda_0$,
$a$, $\Delta$, $\beta$ and initial conditions $\{m_i^0\}$.

There are two candidates of the stopping criterion.
A possible criterion is
``If, for all $i$, the value of $m_i$ is sufficiently
close to $1$ or $-1$ and does not change in
several consecutive iterations, end up the calculation''.
If we can check at every step 
whether the output is a solution of the problem, we 
can use another criterion
based on the ``forced discretization'' $\widehat{\sigma_i}$
of $m_i$ defined as follows:
\begin{eqnarray}
\widehat{\sigma}_i = & 1 & \mbox{ \ \ \ if \ $m_i>0$,} \nonumber \\
\widehat{\sigma}_i = & -1 & \mbox{ \ \ \ otherwise.}
\end{eqnarray}
With this stopping criterion, we end up the calculation
before reaching prescribed maximum number of the iteration
if and only if the target conformation is
the unique ground state of the polymer with the sequence  
$\widehat{\sigma}_i$.

\section{Related Works}
\label{secRW}

In this section, we discuss some of
earlier works in connection with the design equation approach.
First, the use of the soft spin variables is, in itself, not a new technique for the protein design. They are also introduced as optimization tools in the
refs.~\citen{DelKoe,KoeDel,Kono}.
In these references, however, criteria similar to
that of Shakhnovich and Gutin \cite{SG93} are used and
the MTP criterion is not employed. The design equation is
not appeared in these works.

On the other hand, there have been proposed 
several algorithms based on the MTP criterion, which use
different approximations and computational techniques.
In the original algorithm proposed by Deutsch and Kurosky \cite{DK96}
 (see also Kurosky and Deutsch \cite{KD95}), 
the logarithm of the partition function $Z_\beta$ is
approximated by the first order cumulant expansion and
equally-weighted average over all compact self-avoiding conformations
of the polymer. The resultant expression is optimized
through a simulated annealing procedure.
For the generalized HP model, the high
temperature approximation leads to
the expression apparently similar to the the right hand side of the
design equation.
There is, however, a major difference, because
the right hand side of the design equation eq.~(\ref{learn}) can automatically
incorporate the effect of higher order correlations in the conformation space
as the magnitudes of $\{ |m_i| \}$  are
increased. 
Morrissey and Shakhnovich \cite{Morrissey}  also gives
an algorithm based on  MTP criterion, which 
uses a higher order cumulant expansion of the free energy.
These cumulants are evaluated by a mean field approximation 
in conformation space, so that it still keeps the computational
economy of the algorithms based on the 
Shakhnovich and Gutin's criterion \cite{SG93}.
Its performance seems mostly
dependent on the validity of the approximation in the
conformation space.  We will discuss another aspect 
of their work at the end of the paper. 

Seno {\it et al}. \cite{SVMB96}
developed a dual Monte Carlo algorithm that is most faithful to
the MTP criterion.
In their algorithm the partition function $Z_\beta$
is calculated by an important sampling in the conformation space
and the calculated value of $P(\widetilde{\boldr}| \boldsig )$
is optimized through a simulated annealing procedure in the sequence space.
They also test the performance
of the algorithm where the Monte Carlo calculation of the partition function 
$Z_\beta$ is replaced by the exact enumeration of conformations.
In the next section, we compare the performance of an algorithm 
based on the design equation with this version of algorithm.
A significant difference between our approach and theirs
is that the average $\langle \delta(|r_i-r_j|-1) \rangle_\beta$
over the Gibbs distribution eq.~($\ref{mGibbs}$) 
is required in our approach, instead of
the partition function $Z_\beta$ required
in the algorithms by Seno {\it et al}. 

For any algorithm based on the simulated
annealing in the sequence space, the value of a cost function
should be evaluated once per {\it a spin flip } in the
simulated annealing procedure. For example, in
the algorithms of Seno {\it et al}., the value
of the partition function $Z_\beta$ is calculated
once per a trial of changing the type of one monomer
in the sequence. In the design equation approach,
the canonical averages $\langle \delta(|r_i-r_j|-1) \rangle_\beta$
is calculated only once per {\it an iteration } of
the discretized design equation. As will be shown later,
this difference causes a significant advantage
of the algorithms based on the design equation 
when the computational cost of
the cost function and its derivatives are
comparable and both intensive.

\section{Numerical Experiments}
\label{secS}

Here, we test the design equation approach for a three dimensional generalized HP model (\ref{energy}) on the cubic lattice with 27 monomers. In the experiments, 
we restrict the conformations of the polymer to maximally compact self-avoiding ones filling the $3\times  3\times
3$ lattice.  Numerical experiments with compact conformations on small
lattices are common in the study of protein folding \cite{KS92} and  
inverse folding \cite{SG93,DK96,SVMB96} because then the exact enumeration of the conformations is possible.

In the following experiments, the canonical averages $\langle
\delta(|r_i-r_j|-1)
\rangle_\beta$ are calculated by the exact enumeration. 
The parameters
$(\epsilon_1,\epsilon_2)=(-1,0)$ and $(-2.3,-1)$ are used.
We set $\epsilon_3=0$ throughout the experiments.
Note that if we restrict our attention to the maximally compact 
conformations \cite{foot3}, the result is not affected by the addition of
a common constant to all of the parameters $\epsilon_1, 
\epsilon_2, \epsilon_3$ (equivalently, by the value of the parameter $\epsilon_c$). 
Thus, we can set $\epsilon_3=0$ 
without any loss of generality.
We test both of the modified 
energy  function $U^1$ , $U^2$ defined in
eq.~(\ref{Utwo}) and eq.~(\ref{Uone}).  

In Fig.1, an example of the designing process is shown.
In this case, the energy function $U^1$ is used and the
algorithm successfully found a sequence
that has the target conformation as the unique ground state
(a ``good'' sequence).
\begin{fullfigure}
\leavevmode
\epsfxsize=15cm
\epsfbox{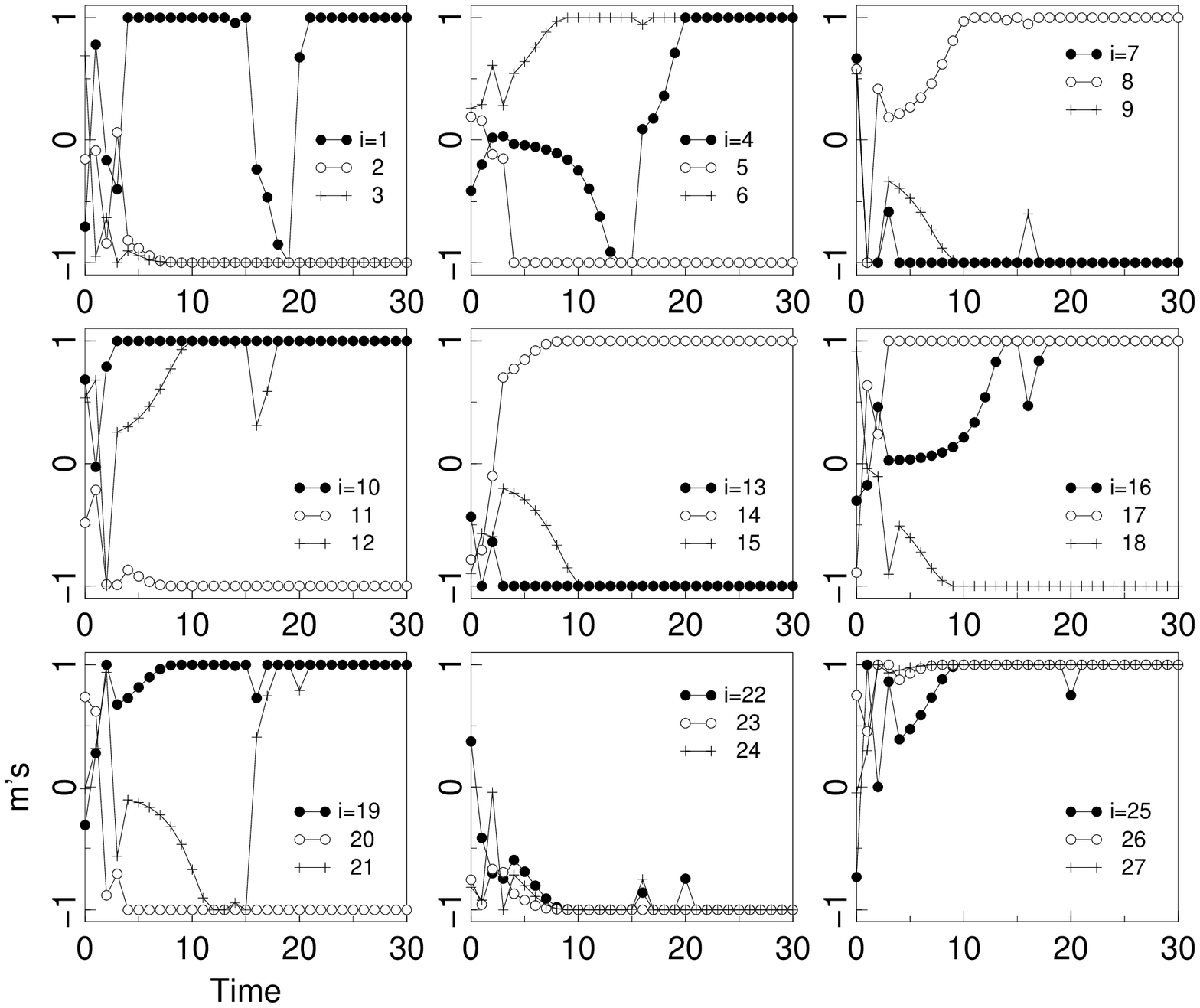}
\caption{
An example of the designing process. \\
The value of each $m_i$ is plotted versus the number of the iteration. A
successful convergence to a good sequence of the target conformation \#1
(see Table~I) is shown.  Note that ``forced discretization'' provides a
good sequence with 20 iterations of the discretized design equation.  Here
$(\epsilon_1, \epsilon_2)=(-1,0)$ and $ \Delta=0.5, \lambda_0=0.0,
a=0.5$.}
\end{fullfigure}

To check the performance of the present algorithm,
we execute the test calculations for five different target  
conformations in Table I. The conformation \#1 is ``the most designable one''
for the generalized HP model on the $3\times 3\times 3$ lattice
according to the definition in ref.\citen{NEC96}. It seems to be one
of the easiest target also in the present cases.
The conformations \#2 - \#5 are randomly chosen ones and some of them
are much more difficult as a target than \#1.  For each target
conformation, 200 trials with different initial sequences $\{m_i^0\}$
are performed.
The number of the successful attempts that result in a ``good''
sequence is recorded.  We also record the number of the iterations
needed to reach a good sequence when we use ``forced dicretization'' as a
stopping criterion.
The initial values $\{m_i^0\}$ of $\{m_i\}$ 
are generated randomly within the
range $[-w,w]$ using a uniform random number.
The temperature $1/\beta$ in MTP criterion 
is set to $0.01$. 
The values of the control parameters in the design equation are
$\Delta=0.5, \lambda_0=0.0, a=0.5$.

To get a feeling of the difficulty of the problem,
we also implement an algorithm based on simulated annealing in the sequence
space. This algorithm, to which we will refer as ``SA'', 
is essentially an algorithm by Seno {\it et al}. 
\cite{SVMB96} with the calculation of $Z_\beta$ 
by the exact enumeration.
In the experiments of SA,
10 trials with different initial sequences are performed
for each target conformation.
The fictitious temperature $T_{SA}$ of the simulated annealing
is initially set as $T_{SA}^0=2.0$
and updated by $T_{SA} := 0.8 \times T_{SA}$
in every Monte Carlo step per spin.
Note that the fictitious temperature
$T_{SA}$, which is introduced for the optimization in the sequence space, is
independent of the temperature $1/\beta$ that determines the Gibbs distribution
eq.~(\ref{Gibbs}) in the conformation space. The value of
$1/\beta$ is kept $0.01$ throughout the experiments.

The results of the experiments
with $(\epsilon_1,\epsilon_2)=(-1,0)$ and $(-2.3,-1)$ are shown in Table~II
and Table~III, respectably.

\begin{table}
\begin{tabular}{cl}
\hline
\#$1$ & $ brf_2l_2u_2b_2r_2f_2ldrb_2l_2dfuru$  \\ \hline
\#$2$ & $ r_2f_2ubufl_2d_2brfubulbr_2dl_2f$  \\ \hline
\#$3$ & $ r_2f_2ublulfd_2brfu_2rb_2dluldf$ \\ \hline
\#$4$ & $ r_2f_2ul_2ub_2druf_2rb_2dfldflbu$  \\ \hline
\#$5$ & $ r_2f_2ubldfu_2rblbrdl_2uf_2d_2bu$ \\
\hline
\end{tabular}
\caption{
The target conformations in the experiments.
Each conformation is encoded by the symbols
 $r$(ight), $l$(eft), $u$(p), $d$(own), $f$(forward), $b$(ackward).
The expressions $r_2$, $f_2$ ..  are
contracted forms of $rr$,$ff$ etc.}
\end{table}

\begin{fulltable}

  \begin{fulltabular}{@{\hspace{\tabcolsep}\extracolsep{\fill}}cccccccccc} \hline
  Target & \multicolumn{3}{c}{$U^1$} &
  \multicolumn{3}{c}{$U^2$} &
  \multicolumn{3}{c}{SA} \\ \cline{2-10}
   & $R$ & $S$ & $P$ & $R$ & $S$ & $P$ & $R$ & $S$ & $P$ \\ \hline
  \#1 &
  0.82\numast & 12.2 & \mbox{ 6.7} &
  0.56\numdag & 16.4 & \mbox{ 3.4} &
  1.0 & 162.0 & \mbox{ 0.62}  \\ \hline
  \#2 &
  0.215\numast & 15.0 & \mbox{ 1.4} &
  0.16\numast & 14.8 & \mbox{ 1.1} &
  0.9 & 294.3 & \mbox{ 0.31}\\
   &
  0.20\numdag & 18.5 & \mbox{ 1.1} &
  0.25\numdag & 14.0 & \mbox{ 1.8} & & &\\ \hline
  \#3 &
  0.30\numast & 16.0 & \mbox{ 1.9} &
  0.355\numdag & 15.9 & \mbox{ 2.2} &
  0.8 & 162.0 & \mbox{ 0.49}\\ \hline
  \#4 &
  0.04\numast & 17.6 & \mbox{ 0.23}&
  0.055\numdag & 14.4 & \mbox{ 0.38}&
  0.4 & 256.5 & \mbox{ 0.16}\\ \hline
  \#5 &
  0.08\numast & 12.1 & \mbox{ 0.66}&
  0.11\numdag & 14.7 & \mbox{ 0.75}&
  0.1 & 232.2 & \mbox{ 0.043}\\ \hline
 \end{fulltabular}
\caption{
The results with $(\epsilon_1,\epsilon_2)=(-1,0)$ for the target
conformations in Table~I.  Rates ($R$) of finding good sequences, average
numbers ($S$) of iterations (or spin flips in SA) and ``efficiency''
$P=100\cdot R/S$ are shown in the cases of (1) the proposed algorithm with
the energy function $U^1$, (2) the proposed algorithm with the energy
function $U^2$, and
(3) an algorithm based on simulated annealing  in the sequence space (SA).
In SA, total number of spin flips until a ``good'' sequence
is used as a correspondence of the number of the iteration
of the design equation, because
the calculation of the partition function
is required in each spin flip in SA.
The initial condition $m_i^0$ is generated randomly 
within $(-w,w)$, where (*) $w=1.0$, (\dag) $w=0.1$.
}
\end{fulltable}

\begin{fulltable}
  \begin{fulltabular}{@{\hspace{\tabcolsep}\extracolsep{\fill}}cccccccccc} \hline
  Target & \multicolumn{3}{c}{$U^1$} &
  \multicolumn{3}{c}{$U^2$} &
  \multicolumn{3}{c}{SA} \\ \cline{2-10}
   & $R$ & $S$ & $P$ & $R$ & $S$ & $P$ & $R$ & $S$ & $P$ \\ \hline
  \#1 &
  1.0\numast & 8.5 & \mbox{ 11.8}&
  0.75\numdag & 10.0 & \mbox{ 7.5}&
  1.0 & 105.3 & \mbox{ 0.95}  \\ \hline
  \#2 &
  0.005\numast & 1.0 & \mbox{ 0.5} &
  0.005\numast & 12.0 & \mbox{ 0.04}&
  0.9 & 186.3 & \mbox{ 0.48}\\
   &
  0.01\numdag & 2.5 & \mbox{ 0.4} &
  0.1\numdag & 10.8 & \mbox{ 0.93} & & &\\ \hline
  \#3 &
  0.002\numast & 5.0 & \mbox{ 0.04} &
  0.225\numdag & 13.0 &  \mbox{ 1.7}&
  0.8 & 210.6 & \mbox{ 0.38}\\ \hline
  \#4 &
  0.00\numast & --- & --- &
  0.015\numdag & 4.8 & \mbox{ 0.31} &
  0.0 & --- & --- \\ \hline
  \#5 &
  0.00\numast & --- & --- &
  0.045\numdag & 11.7 & \mbox{ 0.38}&
  0.3 & 251.3 & \mbox{0.12}\\ \hline
  \end{fulltabular}

\caption{
The results with $(\epsilon_1,\epsilon_2)=(-2.3,-1)$ .
The set of the targets  is the same as that in the Table~II (see Table~I).
The meaning of the notations are shown in the caption
of Table~II.}
\end{fulltable}

The design equation with the modified energy function $U^2$ successfully
finds at least one good sequence for each of the five targets both for
$(\epsilon_1,\epsilon_2)=(-1,0)$ and $(-2.3,-1)$.  The comparison to
SA shows that the cost $S$ of the calculation in each successful run is much less 
in the proposed algorithm than in SA.
This result is highly dependent on the computational
advantage of the design equation approach, i.e, 
the economy of the evaluation of the cost function
that we have already mentioned at the end of 
the previous section.
Note that this advantage will be larger as the length of the
polymer increases.
On the other hand, the rate $R$ of the success is lower in
the proposed algorithm in most cases. Then overall performance seems
comparable in both algorithms. If we define an index $P=100 \cdot R/S$ as a
measure of efficiency, the proposed algorithm
with $U^2$ makes better scores than SA in most cases.

Our experiments show that the performance with the energy function
$U^1$ is rather poor in the case of
$(\epsilon_1,\epsilon_2)=(-2.3,-1)$.
On the other hand, the algorithm works well with the energy function
$U^2$ in both cases $(\epsilon_1,\epsilon_2)=(-1,0)$ and $(-2.3,-1)$.  A
possible reason of this difference is that the energy $U^2$ is a
better representation of the original energy function eq.~(\ref{U1}) in
the sense that the factor $m_i m_j$ in the first term and the factor
$m_i|m_i|+m_j|m_j|$ in the second term of eq.~(\ref{Utwo}) are both 
in the same order of magnitude for small values of
$|m_i|$s.  In the energy function $U^1$, the second term
$\frac{\epsilon_b}{2}(m_i+m_j)$ in eq.~(\ref{Uone}) dominates when the  
values of $|m_i|$s are small and the early stage of the designing process is mostly driven by the second term when we start from an initial condition with
small $|m_i|$s.

\section{Summary and Future Problems}

In Summary, we have proposed a novel approach for heteropolymer design that is
based on the Maximum Target Probability criterion by
Kurosky and Deutsch. The essential point of the proposed approach is the introduction of soft spins (fuzzy monomers) 
that leads to a {\it design equation}, which
is an analog of the Boltzmann machine learning equation in heteropolymer design.
We have tested this approach, which we call  {\it the design equation approach},
for a generalized HP model on the $3\times 3\times 3$ lattice
and have shown that it could successfully find good sequences for target conformations with different degrees of the difficulty.  
With these examples,  our implementation of  the design equation approach 
shows at least comparable performance to an algorithm based on simulated 
annealing in sequence space, when a suitable form of the modified energy function $U^*$ is chosen. 

In this paper,  we have evaluated the performace of the algorithm by the ability to
give a ``good sequence'',  a sequence whose unique ground state coincides with a 
desired connformation. As has been mentioned in the introduction of the paper, it is also
interesting to test dynamical property of the outputs of the algorithm below
their folding temperatures. On the other hand,
a recent work \cite{Morrissey} of Morrissey and Shakhnovich arises a
problem of selecting sequences 
whose equilibrium state at a finite, not necessarily low, temperature
is dominated by a given shape.
The design equation approach might also be useful in this problem. For this purpose,
however, further study of the behavior of the design equation at finite
temperature $1/\beta$ is required.

The design equation approach is fairly general and various modifications are
possible. Here we touch on a few important extensions. First, we can
substitute a Monte Carlo simulation for the exact enumeration in the present
algorithm. This is important because exact enumeration of the conformations 
is impossible for longer or off-lattice polymers. 
Research in this direction is
now in progress and we can successfully implement an algorithm 
based on the design equation with a dynamical Monte Carlo simulation.
The results will be reported in the forthcoming paper
\cite{IKT98}.

Another important challenge is the extension to the cases with
monomers of many different types (many letter cases). Although a
formal extension to many letter cases is not difficult, the test of
the performance of the design equation in such cases is also left for the
future study.

\acknowledgements
This work was supported by a Grant-in-Aid from the Ministry of
Education Science and Culture of Japan, Japan Society for the
Promotion of Science. A part of the computation in this work has been done using
the facilities of the Supercomputer Center, Institute for Solid State
Physics, University of Tokyo.


\begin{thebibliography}{9999}
\bibitem{foot1}
In fact, Kurosky and Deutsch proposed a more
general framework in refs.~\citen{KD95} and \citen{DK96}.  
We, however, do not discuss it further here.
\bibitem{foot3}
In a sense, it corresponds to the situation where
the constant $\epsilon_c$
has a large positive value, which causes compactification
of the polymer.
\end{thebibliography}
\end{document}